\documentclass[12pt]{iopart}

\usepackage{iopams}  
\usepackage{graphicx}
\usepackage{subfig}
\usepackage{xcolor}
\usepackage{ulem}
\usepackage{adjustbox}
\usepackage{pdflscape}

\usepackage{multirow}

\newcommand{\bq}{\begin{equation}}
\newcommand{\eq}{\end{equation}}
\newcommand{\bqn}{\begin{eqnarray}}
\newcommand{\eqn}{\end{eqnarray}}
\newcommand{\nb}{\nonumber}
\newcommand{\lb}{\label}

\begin{document}

\title{Is Birkhoff's Theorem Valid in Einstein-Aether Theory?}

\author{R. Chan$^{1}$, M. F. A. da Silva$^{2}$ and V. H. Satheeshkumar$^{3}$} 

\address{	
			$^{1}$Coordena\c{c}\~{a}o de Astronomia e Astrof\'{i}sica,
			Observat\'{o}rio Nacional (ON), Rio de Janeiro, RJ 20921-400, Brazil
			\\
			$^{2}$Departamento de F\'{i}sica Te\'{o}rica, 
			Universidade do Estado do Rio de Janeiro (UERJ), Rio de Janeiro, RJ 20550-900, Brazil
			\\
			$^{3}$Departamento de F\'{\i}sica, Universidade Federal do Estado do Rio de Janeiro (UNIRIO), Rio de Janeiro, RJ 22290-240, Brazil
		}
			
\ead{chan@on.br, mfasnic@gmail.com, vhsatheeshkumar@gmail.com}

\date{\today}

\begin{abstract}
We attempt to answer whether Birkhoff's theorem (BT) is valid in the Einstein-Aether (EA) theory. The BT states that any spherically symmetric solution of the vacuum field equations must be static, unique, and asymptotically flat. For a general spherically symmetric metric with metric functions $A(r,t)$ \& $B(r,t)$, and aether components $a(r,t)$ \& $b(r,t)$, we prove the conditions for the staticity of spacetime using two different methods. We point out that BT is valid in EA theory only for special values of $c_1+c_3$, $c_1+c_4$, and $c_2$, where we can show that all these special cases are asymptotically flat. In particular, when the aether has only a temporal component, i.e., $b(r,t)=0$ and the $c_{14} \neq 0$ case gives us spherically symmetric static black holes without horizons; that is, they have naked singularities, at least for special values of $c_{14}$. Thus, the cosmic censorship conjecture is violated for the case BT holds. However, when we have an aether vector with temporal and radial components, we only prove that the staticity and the flatness at infinity hold for a special metric and particular combination of the aether parameters. For this case, there exist universal horizons instead of naked singularities.
\end{abstract}

\maketitle

\section{Introduction}

In 1923, mathematician George Birkhoff proved a fundamental theorem in his textbook \textit{Relativity and Modern Physics} \cite{Birkhoff} according to which the Schwarzschild spacetime is the unique spherically symmetric solution of the vacuum Einstein field equations. It implies the absence of time-dependent and spherically symmetric vacuum solutions. That is to say that the vacuum spherically symmetric solutions are static and independent of variations in the matter distribution sourcing the gravitational field as long as the spherical symmetry is preserved. Jebsen independently discovered this theorem \cite{Jebsen} two years earlier. See \cite{VojeJohansen:2005nd} for a fascinating history of this discovery. The Birkhoff theorem (BT) requires that a spherically symmetric solution must be static, given by the Schwarzschild solution. Furthermore, it also implies the absence of gravitational radiation for pulsating or collapsing spherically symmetric bodies. That means there exists no zero-mode of the gravitational wave; correspondingly, in quantum mechanical terms, we can say that zero-mode of graviton does not exist \cite{Weinberg1972}. This reiterates the fact that, in General Relativity (GR), the lowest multipolar gravitational radiation that propagates is quadrupole radiation. 

Because of the experimentally verifiable strict constraints, the BT is of great interest in distinguishing alternative theories of gravity from GR. This theorem has been studied in a class of higher curvature theories, namely the Lovelock theories \cite{Zegers:2005vx, Oliva:2011xu, Oliva:2012zs}. It was proved that the theorem is valid for the Ho\v{r}ava-Lifshitz (HL)  theory which admits the GR limit in the low-energy (IR) region, while the theorem can be violated in the high-energy (UV) region due to nonlinear effects \cite{Devecioglu:2018ote}. As Einstein-Aether (EA) is closely related to HL, it is pertinent to investigate this theorem. 

The EA theory offers a way to study the effects of Lorentz violation in a gravitational sector. The Lorentz invariance (LI) is an exact symmetry in experimentally verified theories of the standard model of particle physics, while in GR, it is only a local symmetry in freely falling inertial frames  \cite{Moore:2013sra}. The violation of LI in the gravitational sector is less well explored than in matter interactions, where several precision experiments highly constrain it \cite{Bars:2019lek}. Besides, in some cases, breaking LI allows us to construct a mathematically consistent quantum gravity \cite{Li:2015itk}. Jacobson and his collaborators introduced and analyzed a general class of vector-tensor theories called the EA theory \cite{Jacobson:2000xp} \cite{Eling:2003rd} \cite{Jacobson:2004ts} \cite{Eling:2004dk} \cite{Foster:2005dk} to study the effects of violation of LI in gravity. Thus, it is of fundamental importance to check whether BT is still valid or needs to be modified for theories that break LI. A brief review of the vector-tensor theories of gravity can be found in \cite{Satheeshkumar:2021zvl}. The first spherical static vacuum solutions in the EA theory were obtained by Eling and Jacobson in 2006 \cite{Eling2006}. Since then, several more solutions have been found, including our recent analytical solutions for static aether \cite{Chan2020}. Most of the literature on black holes in EA theory can be found in the papers \cite{Eling2007}-\cite{Adam2021}.

There are two ways to prove the staticity of a given spacetime which is necessary to prove the BT. The first one uses the results of the reference \cite{Barnes1973}, based on the existence of a fourth timelike Killing vector. If spacetime has only spacelike Killing vectors, then the staticity property is not obeyed. If spacetime has a timelike Killing vector besides the three spacelike Killing vectors, then the metric staticity condition is obeyed. The second one is solving the field equations with the most general spherically symmetric time-dependent metric and showing that the solution is a static spacetime. This is the method commonly followed in GR \cite{Weinberg1972}. Notice that the first method does not require obtaining solutions to the field equations to analyze the existence of the staticity condition. These two methods are necessary but not sufficient conditions to prove the BT. We still have to prove that these solutions are all asymptotically flat and unique. In this work, we use both methods in order to compare the results whenever it is possible. 

In this article, we prove that BT is valid in a general case in EA theory where the aether vector has two components, time and radial components, for a special metric; and it is valid when the aether vector has only a time component. For the latter case, there are two possibilities from the EA field equations: $\dot B \neq 0$ or $\dot B = 0$. We have shown that the case $\dot B \neq 0$ does not provide a solution for the field equations. On the other hand, in the case $\dot B = 0$ we have two possibilities, which are $c_{14} = 0$ and  $c_{14} \neq 0$. We only consider $c_{14} \neq 0$ since EA theory reduces to GR in the case $c_{14}=0$ and guarantees the existence of BT. For the found solutions, the metric function $g_{tt}$ also depends on an arbitrary function of the time, which can be eliminated by redefining a new time coordinate. 
While the $c_{14} \neq 0$ case gives a solution of static black holes in EA theory characterized by one parameter, the mass when the constant $c_{14}$ is fixed.

The paper is organized as follows. Section $2$ briefly outlines the EA theory, whose field equations are solved for a general spherically symmetric metric in Section $3$. In Section $4$, we show using Killing vectors that the only way to have BT valid is to have the aether vector with only a time component. In Section $5$, we study the case $\dot B \neq 0$. In Section $6$, we study the case $\dot B = 0$; in Subsections $6.1$ to $6.2$, we present two solutions of the field equations. In Section 7, we prove the staticity and flatness at infinity of the spacetime for an aether vector with space and time components. We summarize our results in Section $8$.

\section{Field equations in the EA theory}

Here we give a lightning review of the EA theory for completeness. This section has been presented with slight modifications in our previous articles \cite{Campista:2018gfi, Chan:2019mdn, Chan2020, Chan2022}. The general action of the EA theory is given by  
\bq 
S = \int \sqrt{-g}~(L_{\rm Einstein}+L_{\rm aether}+L_{\rm matter}) d^{4}x,
\label{action}
\eq
where the first term is the usual Einstein-Hilbert Lagrangian, defined by $R$, the Ricci scalar, and $G$, the EA gravitational constant, as
\bq 
L_{\rm Einstein} =  \frac{1}{16\pi G} R. 
\eq
The second term, the aether Lagrangian is given by
\bq 
L_{\rm aether} =  \frac{1}{16\pi G} [-K^{ab}{}_{mn} \nabla_a u^m
\nabla_b u^n +
\lambda(g_{ab}u^a u^b + 1)],
\lb{LEAG}
\eq
where the tensor ${K^{ab}}_{mn}$ is defined as
\bq 
{K^{ab}}_{mn} = c_1 g^{ab}g_{mn}+c_2\delta^{a}_{m} \delta^{b}_{n}
+c_3\delta^{a}_{n}\delta^{b}_{m}-c_4u^a u^b g_{mn},
\lb{Kab}
\eq
being the $c_i$ dimensionless coupling constants, and $\lambda$ a Lagrange multiplier enforcing the unit timelike constraint on the aether, and 
\bq
\delta^a_m \delta^b_n =g^{a\alpha}g_{\alpha m} g^{b\beta}g_{\beta n}.
\eq
Finally, the last term, $L_{\rm matter}$, is the matter Lagrangian, which depends on the metric tensor and the matter field.

In the weak-field, slow-motion limit EA theory reduces to Newtonian gravity with a value of  Newton's constant $G_{\rm N}$ related to the parameter $G$ in the action (\ref{action}) by  \cite{Garfinkle2007},
\bq
G = G_N\left(1-\frac{c_{14}}{2}\right).
\lb{Ge}
\eq
Here, the constant $c_{14}$ is defined as
\bq
c_{14}=c_1+c_4.
\lb{beta}
\eq

The field equations are obtained by extremizing the action with respect to independent variables of the system. The variation of the action with respect to the Lagrange multiplier $\lambda$ imposes the condition that $u^a$ is a unit timelike vector, thus 
\bq
g_{ab}u^a u^b = -1,
\label{LagMul}
\eq
while the variation of the action with respect $u^a$, leads to \cite{Garfinkle2007}
\bq
\nabla_a J^a_b + c_4 u^b \nabla_b u_a \nabla_b u^a + \lambda u_b = 0,
\lb{lamu}
\eq
where,
\bq
J^a_m=K^{ab}_{mn} \nabla_b u^n,
\eq
and
\bq
a_a=u^b \nabla_b u_a.
\lb{A}
\eq

The variation of the action with respect to the metric $g_{mn}$ gives the dynamical equations,
\bq
G^{Einstein}_{ab} = T^{aether}_{ab} +8 \pi G  T^{matter}_{ab},
\label{EA}
\eq
where
\bqn
G^{Einstein}_{ab} &=& R_{ab} - \frac{1}{2} g_{ab} R, \nb \\
T^{aether}_{ab}&=& \nabla_c [ J^c\;_{(a} u_{b)} + u^c J_{(ab)} - J_{(a} \;^c u_{b)}] \nb \\
& &- \frac{1}{2} g_{ab} J^c\;_d \nabla_c u^d+ \lambda u_a u_b  \nb \\
& & + c_1 [\nabla_a u_c \nabla_b u^c - \nabla^c u_a \nabla_c u_b] + c_4 a_a a_b, \nb \\
G^{EA}_{ab}&=&G^{Einstein}_{ab} - T^{aether}_{ab}, \\
T^{matter}_{ab} &=&  \frac{- 2}{\sqrt{-g}} \frac{\delta \left( \sqrt{-g} L_{matter} \right)}{\delta g_{ab}}.
\label{fieldeqs}
\eqn

Later, when we solve the field equations (\ref{EA}), we do take into consideration the equations (\ref{LagMul}) in the process of simplification. Thus, in this paper (as in the equations (\ref{Gtt})-(\ref{Gthetatheta}) below), we seem to solve only the dynamical equations, but in fact, we are also solving the equations arising from the variations of the action with respect $\lambda$ and $u^a$. In a more general situation, the Lagrangian of GR is recovered if and only if the coupling constants $c_i$ are identically zero, e.g., considering the equations (\ref{Kab}) and (\ref{LagMul}).

\section{Proof of metric staticity using Killing vectors}

We start with the most general spherically symmetric metric
\bq
ds^2= -A(r,t)\, dt^2 + B(r,t)\, dr^2 + C(r,t)\, dt\, dr+ D(r,t)\, (d\theta^2 + \sin^2 \theta \, d\phi^2).
\lb{ds2g}
\eq
We can simplify this metric further by using the freedom in the choice of coordinates. Transforming to a new radial coordinate, we can set $D(t, r)= r^2$ and remove the cross term by transforming it to a new time coordinate. See Weinberg (1972) \cite{Weinberg1972} for more details. This reduces the form of the line element to,
\bq
ds^2= - A(r,t)\, dt^2 + B(r,t)\, dr^2 + r^2 \, (d\theta^2 + \sin^2 \theta \, d\phi^2).
\lb{metric}
\eq
Let us choose first the aether vector of the form,
\bq
u^a=\left[ a(r,t),b(r,t),0,0,0\right].
\eq
Imposing condition equation (\ref{LagMul}), we can eliminate $a(r,t)$,
\bq
u^a = \left[ \frac{\sqrt{A(r,t) [b(r,t)^2 B(r,t)+1] }}{A(r,t)}, b(r,t), 0, 0\right].
\lb{uar}
\eq
The rationale behind this choice of the aether vector is that it lets us have regular black holes by letting the Killing vector become null on the horizon instead of staying timelike \cite{Eling2006}.

In 1973,  Barnes \cite{Barnes1973} proved that the staticity condition is valid if a spherically symmetrical vacuum spacetime admits a fourth hypersurface orthogonal timelike Killing vector. This is a necessary condition but insufficient because we must also show that the metric is asymptotically flat. However, the asymptotic flatness condition can be proved only if we first resolve the field equations, which cannot be achievable in some cases. That is why we focus on some particular values of $c_2$, $c_1+c_3$, and $c_1+c_4$, where we can solve the field equations analytically.

Although the form of the metric, being a geometrical entity, is the same in all theories of gravity, the metric coefficients $A$ and $B$ have different forms because the field equations of the given theory constrain them. We found that there exist three spacelike Killing vectors for  arbitrary metric coefficients $A(r,t)$ and $B(r,t)$
using the metric (\ref{metric}), that is
\bqn
\xi^1 &=& \sin(\phi) \, \partial_\theta + \cot(\theta)\, \cos(\phi) \, \partial_\phi,\nb\\
\xi^2 &=& \cos(\phi) \, \partial_\theta - \cot(\theta)\, \sin(\phi) \, \partial_\phi,\nb\\
\xi^3 &=& \partial_\phi.
\lb{Kill}
\eqn
A quick calculation shows that these three Killing vectors obey the Lie algebra of the SO(3) group,
\bqn
\left[ \xi^1, \xi^2 \right] &=& \xi^3,\nb\\
\left[ \xi^2, \xi^3 \right] &=& \xi^1,\nb\\
\left[ \xi^3, \xi^1 \right] &=& \xi^2.
\lb{SO3}
\eqn
This implies that the metric posses spherical symmetry.

The timelike Killing vector exists only for $B(r,t)=B(r)$ and for either $A(r,t)=A(r)$ or when $A(r,t) = A_1(r) A_2(t)$. See the timelike Killing vector in the Table \ref{table1}, the Cases $(vi)$, $(vii)$, $(xi)$ and $(xii)$, for details. In this Table, we have introduced the aether vector into the metric using the equation (\ref{LagMul})  without solving the field equations. 

For the spherical symmetry, the metric can have arbitrary $A(r,t)$ and $B(r,t)$. Following, we summarize all the possibilities of the metric and the corresponding Killing vectors for each case. The spacelike Killing vectors are the same equations (\ref{Kill}) when $A(r,t)$ and $B(r,t)$ are assumed to have a variety of functional forms.	Whereas the staticity is obeyed only when $B(r,t) = B_1(r), $ and $A$ are either purely radial [Cases $(vii)$ and $(xii)$] or can be separated into radial and temporal parts, so that this separation of variables allows us to have a coordinate transformation such that we can write $A$ as independent of time [Cases $(vi)$ and $(xi)$]. The case $(xi)$ is studied in more detail in Section 6, and the case $(vi)$ in Section 7.

Note that we have studied the case $(vii)$ in an earlier paper and found that BT is satisfied, although only for some values of $c_{14}$ \cite{Chan2022}. This agrees with Eling and Jacobson's claim \cite{Eling2006}\cite{Eling2007} that radial tilting of the aether vector may prevent the spherical solutions from being time-independent and consequently forbid the BT.

In other words, the metric function $A$ cannot have its separation of variables because of the radial component of the aether. (This property will be more explicit in the next section when we solve the field equations for the aether with time component only.)   This fact can be easily seen in the first column of Table \ref{table1} for the Cases $(i)$ to $(v)$. The condition that $u^a$ is a timelike unit vector does not allow this separation. However, when we have only the aether time component, this prevention does not exist anymore [Cases $(viii)$ to $(xii)$].
Interestingly, the well-known fact that a general spherically symmetric metric, by appropriate choice of the time coordinate, can produce the spherically symmetric and static case was already appreciated in 1921 Jebsen's derivation of BT. However, this redefinition cannot be done at will and should be permitted by the theory of gravity under consideration. 
In the next section, we will prove the staticity property of the metric for the particular case of an aether vector with only a time component.

\clearpage

\begin{table*}
	\centering
	\begin{minipage}{150 mm}
		\caption{Summary of all possibilities of the metric and the number of Killing vectors}
		\label{table1}
		\begin{tabular}{@{}|c|c|c|c|c|c|c|}
			\hline
			\multirow{3}{*}{Case} &  &  &  &  & Timelike & Notes\\
			& $A(r,t)$ & $B(r,t)$ & $a(r,t)$ & $b(r,t)$ & Killing & \\
			& & & & & Vector & \\
			\hline
			\multirow{2}{*}{$(i)$} & ${\frac { b \left( r,t \right) ^{2}B \left( r,t
					\right) +1}{ a \left( r,t \right) ^{2}}}$ & $B(r,t)$ & $a(r,t)$ & $b(r,t)$ &(*)&\\
			& & & & & & \\
			\hline
			\multirow{2}{*}{$(ii)$} & $\frac { b_1(r)^2 \, b_2(t)^{2} B_1(r) \, B_2(t)+1}{a_1(r)^2 \, a_2(t)^2}$ & $B_1(r) \, B_2(t)$ & $a_1(r) \, a_2(t)$ & $b_1(r) \, b_2(t)$ &(*)&\\
			& & & & & & \\
			\hline
			\multirow{2}{*}{$(iii)$} & $\frac { b_1(r)^2B_2(t)^{2} B_1(r) \, B_2(t)+1}{a_1(r)^2 \, B_2(t)^2}$ & $B_1(r) \, B_2(t)$ & $a_1(r) \, B_2(t)$ & $b_1(r) \, B_2(t)$ &(*)&\\
			& $B_2(t))=a_2(t)=b_2(t)$& & & & & \\
			\hline
			\multirow{2}{*}{$(iv)$} & $\frac { b_1(r)^2b_2(t)^{2} B_1(r)+1}{a_1(r)^2\,a_2(t)^2}$ & $B_1(r)$ & $a_1(r) \, a_2(t)$ & $b_1(r) \, b_2(t)$ &(*)&\\
			& & & & & & \\
			\hline
			\multirow{2}{*}{$(v)$} & $\frac { b_1(r)^2b_2(t)^{2} B_1(r)+1}{a_1(r)^2\,a_2(t)^2}$ & $B_1(r)$ & $a_1(r) \, a_2(t)$ & $b_1(r) \, a_2(t)$ &(*)&\\
			& $a_2(t)=b_2(t)$ & & & & & \\
			\hline
			\multirow{2}{*}{$(vi)$} &   {$\frac { b_1(r)^2 B_1(r)+1}{a_1(r)^2\,a_2(t)^2}$} &  {$B_1(r)$} & $a_1(r) \, a_2(t)$ & $b_1(r)$ & $-a_2(t)\partial_t$ & (\#) \\
			& & & & & & \\
			\hline
			\multirow{2}{*}{$(vii)$} & ${\frac { b_1 \left( r\right) ^{2}B_1 \left( r \right) +1}{ a_1 \left( r\right) ^{2}}}$ & $B_1(r)$ & $a_1(r)$ & $b_1(r)$ & $-\partial_t$ & (1)\\
			& & & & & & \\
			\hline
			\multirow{2}{*}{$(viii)$} &  $\frac{1}{a(r,t)^2}$ & $B(r,t)$ & $a(r,t)$ & 0 &(*)&\\
			& & & & & & \\
			\hline
			\multirow{2}{*}{$(ix)$} & $\frac{1}{a_1(r)^2 \, a_2(t)^2}$ & $B_1(r) \, B_2(t)$ & $a_1(r) \, a_2(t)$ & 0 &(*)& \\
			& & & & & & \\
			\hline
			\multirow{2}{*}{$(x)$} & $\frac{1}{a_1(r)^2\,B_2(t)^2}$ & $B_1(r) \, B_2(t)$ & $a_1(r) \, B_2(t)$ & 0 &(*)& \\
			& $B_2(t)=a_2(t)$ & & & & & \\
			\hline
			\multirow{2}{*}{$(xi)$} &  $\frac{1}{a_1(r)^2a_2(t)^2}$ & $B_1(r)$ & $a_1(r) \, a_2(t)$ & 0 & $-a_2(t)\partial_t$ & (2)\\
			& & & & & & \\
			\hline
			\multirow{2}{*}{$(xii)$} & $\frac{1}{a_1(r)^2}$  & $B_1(r)$ & $a_1(r)$ & 0 & $-\partial_t$ & (3)\\
			& & & & & & \\
			\hline
		\end{tabular}
		
		\medskip
		
		where $B_1(r)$, $B_2(t)$, $a_1(r)$, $a_2(t)$, $b_1(r)$ and $b_2(t)$ are arbitrary functions of the coordinates $r$ or $t$.  Entries with (*) indicate that only the spacelike Killing vectors exist in these cases and are given by the equations (\ref{Kill}). The detailed solutions of black holes (1), (2), and (3) are given in references \cite{Chan2020}, this work, and \cite{Chan2022}, respectively. (\#) This case is rather unrealistic since both aether components, in general, should have the same time dependence, as in the case $(v)$. However, this case also exemplifies one way that we can have the fourth time Killing vector, i.e.,
		the possibility of the variable separation of $A(r,t)$ as $[b_1(r)^2 B_1(r)+1]/a_1(r)^2\, \times\, 1/a_2(t)^2$, as in the Case $(xi)$.
		
	\end{minipage}
\end{table*}

\section{Spherical Solutions of EA field equations of a aether with time component}

As demonstrated in the previous section, we consider the spherically symmetric metric without any loss of generality,
\bq
ds^2= -A(r,t)\, dt^2 + B(r,t)\, dr^2 + r^2 (d\theta^2 + \sin^2 \theta d\phi^2).
\lb{ds2}
\eq

We now choose to work with the aether field with only a time component,
\bq\,
u^a = \left[ a(r,t), 0, 0, 0\right],
\eq
and when we apply the timelike unitary condition in equation (\ref{LagMul}), we get
\bq
u^a=\left[\frac{1}{\sqrt{A(r,t)}},0,0,0\right].
\lb{uat}\,
\eq

From the equation (\ref{lamu}), we can show that it has only two non-zero components, temporal and radial, respectively are given by
\bqn 
&&- {\frac { c_{13} {\dot B}^{2} r A }{ 4 A^{3/2} B^{2}r}}  
+ {\frac {  ( 2  {\ddot B} B r A -  {\dot A} {\dot B} B  r - 2 {\dot B}^{2} r A ) c_{2}}{4 A^{3/2}  B^{2} r}} + \lambda\, {\frac {4 A^{2} B^{2} r }{ 4 A^{3/2}  B^{2} r}}  \nb \\ 
&&+ {\frac { c_{3} (2 A'' B A \,r - A' B' A r - A'^{2} B r + 4  A' A  B) }{ 4 A^{3/2}  B^{2} r}} + {\frac {2 c_{14} A'^{2} B r }{ 4 A^{3/2}  B^{2} r}} = 0.
\label{u(t)}
\eqn
\bqn 
&&{\frac { c_{13} ( 2 {\dot B'}   A^{2} B r - 2  B' {\dot B} A^{2} r - {\dot B} A B A' r +4 {\dot B} A^{2} B  )}{4 A^{5/2} B^{2}r}} 
\nb \\
&&+ {\frac { c_{2} ( 2 {\dot B'} A^{2} B r - 2 B' {\dot B} A^{2} r - {\dot B} A B A' r ) }{4 A^{5/2} B^{2} r}} \nb \\
&&	+ {\frac { c_{14} (2 {\dot A} B^{2} A' r - 2 {\dot A'} A B^{2} r  + {\dot B} A B A' r)  }{4 A^{5/2} B^{2} r}}=0
\label{u(r)}
\eqn
The EA field equations (\ref{fieldeqs}) for the metric (\ref{ds2}) are given by,
\bqn 
G^{EA}_{tt} =&& 
{\frac { c_{3} ( 4 A'' A B {r}^{2} - 2 B' A A' {r}^{2} + 8 A' B A r ) }{8 {r}^{2} B^{2}A }} 
- \,{\frac { c_{13} ( {\dot B}^{2} A {r}^{2} + 2 B A'^{2} {r}^{2} ) }{8 {r}^{2} B^{2} A }} \nb \\
&& - \,{\frac { c_{14} ( 4 A'' A  B {r}^{2} - 2 B' A  A' {r}^{2} - 7 B A'^{2} {r}^{2} +8 A' B A r ) }{8 {r}^{2} B^{2} A  }} - \lambda \nb \\
&& - \,{\frac { c_{2} ( -4 {\ddot B} A B {r}^{2} + 2 {\dot A} {\dot B} B {r}^{2} +3 {\dot B}^{2} A  {r}^{2} ) }{8 {r}^{2} B^{2}A}}
+ \,{\frac { B'A r + B^{2} A - B A}{ {r}^{2} B^{2}}}\nb\\
\label{GEA_tt}
\eqn		
\bqn
G^{EA}_{tr} =&& G^{EA}_{rt}= {\frac { c_{14} ( 2 {\dot A} A' B r - 2{\dot A'} A B r +  A  A' {\dot B} r ) }{4 A^{2} B r}}+{\frac {{\dot B} }{B r}}
\label{GEA_rt}
\eqn
\bqn
G^{EA}_{rr} =&& {\frac { -(c_{13} +c_{2}) ( 4 {\ddot B} A B {r}^{2} - 2 {\dot A} {\dot B} B {r}^{2} -3 {\dot B}^{2} A {r}^{2} )}{8 A^{2} {r}^{2} B }} + {\frac { A'^{2} c_{14}}{ 8 A^{2}}} \nb \\ 
&& - {\frac {8 B^{2} A^{2} - 8 A' B A r - 8 B A^{2} }{8 A^{2} {r}^{2} B}}
\label{GEA_rr}
\eqn
\bqn 
G^{EA}_{\theta \theta}=&& {\frac { c_{13} {\dot B}^{2} {r}^{2} }{ 8 A  B^{2}}} - {\frac {c_{14} A'^{2} {r}^{2} }{8 B A^{2}}}
-{\frac { c_{2} r  ( 4 {\ddot B} B r A -2 {\dot A} {\dot B} B r - 3 {\dot B}^{2} r A )}{ 8 B^{2} A^{2}}} \nb \\
&&-\frac {r}{ 8 B^{2} A^{2}} ( -4 A'' B r A + 4 {\ddot B} B r A + 2 B' A' r A - 2 {\dot A} {\dot B} B r \nb \\ 
&&  ~~~~~~~~~~~~~ - 2 {\dot B}^{2} r A + 2 A'^{2} B r + 4 B' A^{2} - 4 A' B A)
\label{GEA_thetatheta}
\eqn
\bqn 
G^{EA}_{\phi \phi} =&& \sin\theta^{2} \, G^{EA}_{\theta \theta}
\label{GEA_phiphi}
\eqn
We can combine the above constraints (\ref{u(t)})-(\ref{u(r)}) and field equations to eliminate the lagrangian multiplier $\lambda$ to get the following five differential equations.
\bqn
&&\frac{ c_{13}}{8B^2} {\dot B}^2  +
\frac{c_{14}}{8r^2 B^2 A} (-4 {A''} B r^2 A-8 A r {A'} B
+3 {A'}^2 B r^2+2 {B'} {A'} r^2 A) \nb\\
&& +\frac{ c_{2}}{8B^2} {\dot B}^2 +\frac{1}{8r^2 B^2 A} (8 r {B'} A^2+8 B^2 A^2-8 B A^2)
=0,
\lb{Gtt}\\
~\nb\\
&&-\frac{c_{13}}{4A B^2 r} (-2 A r {\dot B} {B'}+2 A r {\dot B'} B+4 {\dot B} A B -r A' {\dot B} B)\nb\\
&& -\frac{c_{2}}{4A B^2 r} (-r A' {\dot B} B-2 A r {\dot B} {B'}
+2 A r {\dot B'} B) +\frac{\dot B}{B r}=0,
\lb{Gtr}\\
~\nb\\
&&-\frac{c_{14}}{4B r A^2} (-r {A'} {\dot B} A+2 r {\dot A'} A B-2 r {A'} {\dot A} B)
+\frac{\dot B}{B r}=0,
\lb{Grt}\\
~\nb\\
&&-\frac{c_{13}+ c_{2} }{8A^2 r^2 B} (4 {\ddot B} B r^2 A-3 {\dot B}^2 r^2 A-2 {\dot B} {\dot A} B r^2)
+\frac{c_{14}}{8A^2} {A'}^2 \nb\\ &&
-\frac{1}{8A^2 r^2 B} (8 B^2 A^2-8 B A^2
-8 A r {A'} B)=0,
\lb{Grr}\\
~\nb\\
&&\frac{c_{13}}{8A B^2} r^2 {\dot B}^2-\frac{c_{14}}{8B A^2} r^2 {A'}^2  
-\frac{c_{2}}{8B^2 A^2} r (-3 {\dot B}^2 r A+4 {\ddot B} B r A-2 {\dot B} {\dot A} B r) \nb\\
&&-\frac{1}{8B^2 A^2} r (4 {B'} A^2-4 A' B A-2 {\dot B} {\dot A} B r
+4 {\ddot B} B r A+2 {A'}^2 B r-4 {A''} B r A\nb\\
&&+2 {B'} {A'} r A-2 {\dot B}^2 r A)=0,
\lb{Gthetatheta}
\eqn
where the symbols prime and dot denote the differentiation with respect to $r$ and $t$, respectively. We notice that when $c_{13}=0$, $c_{14}=0$, and $c_2=0$, the EA field equations reduce to their counterparts in GR. If we substitute $A(r,t) \rightarrow e^{2{  A}(r)}$ and $B(r,t) \rightarrow e^{2{  B}(r)}$, we obtain exactly the same field equations of our previous paper \cite{Chan2020}. 

Now, we solve these equations. From equation (\ref{Gtr}) we have
\bqn
-\frac{\dot B}{B} \left[ (c_2+c_{13}) \left( \frac{B'}{2B} + \frac{A'}{4A} \right) +\frac{1}{r}(1-c_{13}) \right]
+(c_2+c_{13}) \frac{\dot B'}{2B}=0.
\lb{Gtr2}
\eqn
There are two possibilities from equation (\ref{Gtr2}): $\dot B \neq 0$ or $\dot B = 0$. We will analyze below both cases in detail. 

\section{{\bf Case (I):} $\bf \dot B \neq 0$}

First, we will assume $\dot B \neq 0$. Solving equation (\ref{Gtr2}), assuming $c_2+c_{13} \neq 0$ and $1-c_{13} \neq 0$, thus
\bq
A=\left[ \frac{\dot B}{B}F_0(t)r^{-\frac{2(1-c_{13})}{c_2+c_{13}}} \right]^2,
\lb{Art0}
\eq
where $F_0(t)$ is an arbitrary integration function of the time.

Substituting equation (\ref{Art0}) into (\ref{Grt}) and solving it for $B$ we obtain
\bq
B=\frac{2.3^\frac{1}{4}}{3} \sqrt{2}  G_0(r) \left[ C_2 (-C_1+t)^3 \right]^\frac{1}{4},
\lb{Brt0}
\eq
where $G_0(r)$ is an arbitrary integration function of the radial coordinate and $C_1$ and $C_2$ are arbitrary integration constants. Substituting $A$ and $B$ found into (\ref{Grt}) again we get
\bq
\frac{3}{4}\,{\frac {{c_2}+{c_{13}}\,{c_{14}}+{c_{13}}-{c_{14}}}{\left( {C_1}-t \right) {c_{14}}\,r}}=0,
\eq
thus, we get an additional relation $c_2 + c_{13} c_{14} + c_{13} - c_{14}=0$.

Substituting equations (\ref{Art0}) and (\ref{Brt0}) into the field equations (\ref{Gtr}) and (\ref{Grt}) we can easily show that, if $c_2+c_{13}c_{14}+c_{13}-c_{14}=0$,
they are satisfied identically. On the other hand, substituting equations (\ref{Art0}) and (\ref{Brt0}) into the field equations (\ref{Gtt}), (\ref{Grr}) and 
(\ref{Gthetatheta}) we can notice that they are not satisfied identically. Thus, we have shown that the field equations do not admit  $\dot B \neq 0$ as a solution.

\section{{\bf Case (II):} $\bf \dot B = 0$}

Now, we will assume $\dot B = 0$. From the field equation (\ref{Gtr2}), let us study the case where
\bq 
\frac{\dot B}{B}=0.
\lb{dBrt}
\eq
Thus, we have
\bq 
B(r,t)=f(r).
\lb{Brt}
\eq

Solving simultaneous equations (\ref{Gtt})-(\ref{Gthetatheta}), using (\ref{Brt}) and  Maple 16, we get six possible different solutions (i) $c_2=0$ and $c_{14}=0$; (ii) $c_{13}=0$ and $c_{14}=0$; (iii) $c_{14}=0$; (iv) $c_2=0$ and $c_{13}=0$; (v) $c_{2} =0$; (vi) $c_{13} =0$.

However, we can notice from field equations (\ref{Gtt})-(\ref{Gthetatheta}) that whenever it appears the term in $c_2$ or $c_{13}$, they also multiply $\dot B$. Thus, in this particular case, we have only to consider $c_{14} \neq 0$.  We get the same GR results when $c_{14}=0$.

Let us now analyze these two possible cases in detail instead of six cases. 

\subsection{{\bf Subcase (A):} $\bf c_{14} = 0$}

Assuming equation (\ref{Brt}) and the field equations
(\ref{Gtt})-(\ref{Gthetatheta}) we get
\bqn
&&\frac{1}{8r^2 f^2 A} (8 r {f'} A^2+8 f^2 A^2-8 f A^2)=0,
\lb{GttA}
\eqn
\bqn
&&-\frac{1}{8A^2 r^2 f} (8 f^2 A^2-8 f A^2-8 A r {A'} f)=0,
\lb{GrrA}
\eqn
\bqn
&&-\frac{1}{8f^2 A^2} r (4 {f'} A^2-4 A' f A+2 {A'}^2 f r-4 {A''} f r A+2 {f'} {A'} r A)=0,
\lb{GthetathetaA}
\eqn
whose solution is given by
\bqn
&&A= h(t) \left(1+\frac{C_0}{r}\right),\nb\\
&&f=\frac{r}{r+C_{0}},
\eqn
where $h(t)$, is an arbitrary integration function and $C_0$  is an arbitrary integration constant. We have chosen $C_0=-2 M$ to resemble the Schwarzschild solution as in the GR where $M$, from now on, is the Schwarzschild mass. Thus, $A(r)$ and $f(r)$ can be rewritten as
\bqn
&&A=F_1(t)\left(1-\frac{2M}{r}\right),\nb\\
&&f=\frac{r}{r-2M}.
\lb{AB1}
\eqn
Substituting this solution into the field equations (\ref{Gtt})-(\ref{Gthetatheta}), we can prove that they are satisfied identically.
Since $h(t)$ is an arbitrary integration function, we can notice that we can eliminate redefining the time coordinate, giving the Schwarzschild static solution. Thus, we have proved that the BT is also valid for EA in this case.

The Kretschmann scalar is given by
\bq
K = \frac{ 48 M^2}{r^6}.
\eq
We can note that $r=0$ is the unique singularity of this spacetime.

\subsection{{\bf Subcase (B):} $\bf c_{14} \neq 0$}

Assuming equation (\ref{Brt}) and the field equations (\ref{Gtt})-(\ref{Gthetatheta}) we get
\bqn
&&\frac{c_{14}}{8r^2 f^2 A} (-4 {A''} f r^2 A-8 A r {A'} f+3 {A'}^2 f r^2 +2 {f'} {A'} r^2 A) \nb\\
&&+\frac{1}{8r^2 f^2 A} (8 r {f'} A^2+8 f^2 A^2
-8 f A^2)=0,
\lb{GttB}\\
\lb{GtrB}\\
&&-\frac{c_{14}}{4f r A^2} (2 r {\dot A'} A f-2 r {A'} {\dot A} f) =0,  
\lb{GrtB}\\
&&\frac{c_{14}}{8A^2} {A'}^2 -\frac{1}{8A^2 r^2 f} (8 f^2 A^2-8 f A^2
-8 A r {A'} f)=0,
\lb{GrrB}\\
&&-\frac{c_{14}}{8f A^2} r^2 {A'}^2 -\frac{1}{8f^2 A^2} r (4 {f'} A^2-4 A' f A\nb\\
&&+2 {A'}^2 f r-4 {A''} f r A+2 {f'} {A'} r A)=0,
\lb{GthetathetaB}
\eqn
whose solution is given by
\bqn
\lb{frb}
f &=& \frac{1}{8A^2} ({A'}^2 r^2 c_{14}+8 A^2+8 A r {A'}),\\
\lb{dArt1b}
{\dot A'} &=&  \frac{{A'}\dot A}{A},\\
\lb{dArt2b}
{A''} &=& \frac{1}{8r A^2} (-{A'}^3 c_{14} r^2-16 {A'} A^2).
\eqn

From equation (\ref{dArt1b}) we can see that
\bq
A(r,t)=g(r) h(t),
\eq
where $g(r)$ and $h(t)$ are arbitrary functions.
Thus, we use this variable separation in order to solve the equations 
(\ref{frb}) and (\ref{dArt2b}).
Thus, the equation (\ref{dArt2b}) reduces to
\bq 
\frac{h}{8r g^2}  (8 {g''} r g^2+{g'}^3 r^2 c_{14}+16 {g'} g^2)=0.
\lb{ht}
\eq

Notice that this equation is the same as our previous paper \cite{Chan2020} assuming $g(r)=e^{2A(r)}$ times an arbitrary function of the time $h(t)$, since we do not have any field equation for its determination. Solving equation (\ref{ht}) we get
\bq
\frac{1}{2} \alpha  g^{-\frac{1}{2}+\frac{1}{4} \alpha}-g^{\frac{1}{2} \alpha} C_{2}+C_{1}=0,
\lb{gr}
\eq
where $C_1$, $C_2$ are arbitrary integration constants and $\alpha=\sqrt{-2 c_{14}+4}$. This last transcendental equation can be solved analytically using particular values for $c_{14}$ ($3/2,\, 16/9,\, 48/25$ and $-16$). These particular values of $c_{14}$ are the same used in \cite{Chan2020}. For the GR case  with $c_{14}=0$ we get the Schwarzschild spacetime $g=1-\frac{2M}{r}$ imposing $C_2=C_1=-\frac{1}{2M}$. Only for these values of $c_{14}$ we can guarantee
the flatness of the spacetime at infinity.

Again, since $h(t)$ is an arbitrary function, we can notice that it can be eliminated by redefining the time coordinate, giving a black hole
static solution. 

In order to prove the existence and uniqueness of this solution mathematically, we will use the methods described in the reference \cite{Trench2001} for nonlinear differential equations. First, we obtain the first derivative $g'$ from equation (\ref{gr}). If both $g$ and $g'$ are continuous in a given interval of $r$, then the equation (\ref{ht}) has a unique solution. Thus,
\bq
g'={\frac {4M g ^{\frac{1}{2}+\frac{1}{4}\alpha}}{r \left( 2r g ^{\frac{1}{2}\alpha}-2M g ^{-\frac{1}{2}+\frac{1}{4}\alpha}+ \alpha M g ^{-\frac{1}{2}+\frac{1}{4}\alpha} \right) }},
\eq
whose denominator is null at
\bq
g_0= \left[ {\frac {2r}{M \left( 2-\alpha \right) }} \right] ^{-\frac{4}{2+\alpha}},
\lb{g}
\eq
where $c_{14} \neq 0$, implying $\alpha \neq 2$. Thus, the solution (\ref{gr}) is unique if only if $g \neq g_0$ for given values of $c_{14}$ and $M$.

We have found a solution that depends on $g$, which is fixed when we choose a given value of $c_{14}$. However, $c_{14}$ is not a constant as the mass is for the Schwarzschild solution, where the radial functional dependence does not change with this constant. Here $c_{14}$ defines the theory of EA, i.e., the radial functional dependence does change with this constant, which observational tests will fix. Once this is fixed, the solution is unique. Thus, we have proved that the BT is also valid for EA for the aether static.

\section{Spherical Solutions of EA field equations of a aether with space and time components}

Let us now prove the staticity of the spacetime using the field equations, as done in Section 4. Since we have shown that the unique metric that may satisfy the staticity condition using the Killing vectors is given by case (vi) of Table \ref{table1}, we have
\bq
ds^2= -\left[ \frac { b_1(r)^2 B_1(r)+1}{a_1(r)^2\,a_2(t)^2}\right]\, dt^2 + B_1(r)\, dr^2 + r^2 (d\theta^2 + \sin^2 \theta d\phi^2).
\lb{ds21}
\eq
We now choose to work with the unitary aether field with space and time components given by
\bq
u^a = \left[ a(r,t), b(r,t), 0, 0\right]=\left[ a_1(r) a_2(t), b_1(r), 0, 0\right],
\lb{uat1}
\eq
The aether field equations, collecting the terms $c_2$, $c_{13}$ and $c_{14}$, are given by
\bqn
G^{aether}_{rr}&&=
-\frac{1}{8{{r}^{2}{A_1}^{2}B_1  \left( {b_1}^{2}B_1+1 \right) }} \times\\\nb
&&[\,( 4 {b_1}^{3}{r}^{2}A_1 {B_1}^{3}A'_1 b'_1+4 b_1 {r}^{2}A_1 {B_1}^{2}b'_1 A'_1+2 {b_1}^{2}{r}^{2}A_1 B_1 A'_1 B'_1+ \\\nb
&& 4 b_1 {r}^{2}{A_1}^{2}B_1 b'_1 B'_1+{r}^{2}{A_1}^{2}B_1 {b_1}^{4}{B'_1}^{2}+8 {A_1}^{2}{B_1}^{3}{b_1}^{4}+{b_1}^{2}{B'_1}^{2}{r}^{2}{A_1}^{2}+\\\nb
&& {r}^{2}{B_1}^{3}{b_1}^{4}{A'_1}^{2}+4 {b_1}^{3}{r}^{2}{A_1}^{2}{B_1}^{2}b'_1 B'_1+{b_1}^{2}{A'_1}^{2}{r}^{2}{B_1}^{2}+\\\nb
&& 4 {b_1}^{2}{r}^{2}{A_1}^{2}{B_1}^{3}{b'_1}^{2}+ 8 {b_1}^{2}{A_1}^{2}{B_1}^{2}+4 {b'_1}^{2}{r}^{2}{A_1}^{2}{B_1}^{2}+\\\nb
&& 2 {b_1}^{4}{r}^{2}A_1 {B_1}^{2}A'_1 B'_1 )\, c_{13}+\\\nb
&&( -{r}^{2}{A_1}^{2}B_1 {b_1}^{4}{B'_1}^{2}-{A'_1}^{2}B_1 {r}^{2}- 4 b_1 {r}^{2}A_1 {B_1}^{2}b'_1 A'_1- \\\nb
&& 4 {b_1}^{2}{r}^{2}{A_1}^{2}{B_1}^{3}{b'_1}^{2}-2 {b_1}^{4}{r}^{2}A_1 {B_1}^{2}A'_1 B'_1-4 {b_1}^{3}{r}^{2}{A_1}^{2}{B_1}^{2}b'_1 B'_1-\\\nb
&& 4 {b_1}^{3}{r}^{2}A_1 {B_1}^{3}A'_1 b'_1-{r}^{2}{B_1}^{3}{b_1}^{4}{A'_1}^{2}-2 {b_1}^{2}{A'_1}^{2}{r}^{2}{B_1}^{2}-\\\nb
&& 2 {b_1}^{2}{r}^{2}A_1 B_1 A'_1 B'_1 )\, c_{14}+\\\nb
&&( 16 b_1 r{A_1}^{2}{B_1}^{2}b'_1+8 {b_1}^{2}r{A_1}^{2}B_1 B'_1+{r}^{2}{A_1}^{2}B_1 {b_1}^{4}{B'_1}^{2}+\\\nb
&& 4 b_1 {r}^{2}{A_1}^{2}B_1 b'_1 B'_1+2 {b_1}^{2}{r}^{2}A_1 B_1 A'_1 B'_1+4 {b_1}^{3}{r}^{2}A_1 {B_1}^{3}A'_1 b'_1+\\\nb
&& 4 b_1 {r}^{2}A_1 {B_1}^{2}b'_1 A'_1+2 {b_1}^{4}{r}^{2}A_1 {B_1}^{2}A'_1 B'_1+16 {A_1}^{2}{B_1}^{3}{b_1}^{4}\\\nb
&& 16 {b_1}^{2}{A_1}^{2}{B_1}^{2}+8 {B_1}^{2}A_1 rA'_1 {b_1}^{2}+8 {b_1}^{4}A_1 {B_1}^{3}A'_1 r+\\\nb
&& 16 r{A_1}^{2}{B_1}^{3}{b_1}^{3}b'_1+8 {b_1}^{4}r{A_1}^{2}{B_1}^{2}B'_1+4 {b_1}^{2}{r}^{2}{A_1}^{2}{B_1}^{3}{b'_1}^{2}+\\\nb
&& {b_1}^{2}{A'_1}^{2}{r}^{2}{B_1}^{2}+{b_1}^{2}{B'_1}^{2}{r}^{2}{A_1}^{2}+{r}^{2}{B_1}^{3}{b_1}^{4}{A'_1}^{2}+4 {b'_1}^{2}{r}^{2}{A_1}^{2}{B_1}^{2}+\\\nb
&& 4 {b_1}^{3}{r}^{2}{A_1}^{2}{B_1}^{2}b'_1 B'_1 )\, c_{2} +\\\nb
&&( 8 {B_1}^{2}{A_1}^{2}-8 B_1 {A_1}^{2}-8 {b_1}^{2}{A_1}^{2}{B_1}^{2}-8 A_1 rA'_1 B_1-\\\nb
&& 8 {B_1}^{2}A_1 rA'_1 {b_1}^{2}+8 {A_1}^{2}{B_1}^{3}{b_1}^{2} )\,],\\
\lb{Gaerr1}
G^{aether}_{\theta\theta}&&=
-\frac{1}{8{B_1}^{2}{A_1}^{2} \left( {b_1}^{2}B_1+1 \right) }\times\\\nb
&&[\,  r ( -4 r{A_1}^{2}{B_1}^{3}{b_1}^{2}{b'_1}^{2}-2 {B_1}^{2}A_1 r{b_1}^{4}B'_1 A'_1-4 A_1 r{B_1}^{3}{b_1}^{3}b'_1 A'_1-\\\nb
&& 4 b_1 r{A_1}^{2}B_1 b'_1 B'_1-r{B_1}^{3}{A'_1}^{2}{b_1}^{4}-4 {B_1}^{2}A_1 rb_1 b'_1 A'_1+4 {B_1}^{2}{A_1}^{2}{b_1}^{4}B'_1-\\\nb
&&4 {b'_1}^{2}r{A_1}^{2}{B_1}^{2}-{b_1}^{2}{B'_1}^{2}r{A_1}^{2}+16 b_1 {A_1}^{2}{B_1}^{2}b'_1+4 {b_1}^{2}{A_1}^{2}B_1 B'_1-\\\nb
&&r{A_1}^{2}B_1 {b_1}^{4}{B'_1}^{2}+4 {B_1}^{2}A_1 A'_1 {b_1}^{2}+4 {b_1}^{4}A_1 {B_1}^{3}A'_1-r{B_1}^{2}{A'_1}^{2}{b_1}^{2}-\\\nb
&&4 r{A_1}^{2}{B_1}^{2}{b_1}^{3}b'_1 B'_1-2 A_1 rB_1 {b_1}^{2}B'_1 A'_1+16 {A_1}^{2}{B_1}^{3}{b_1}^{3}b'_1 )\, c_{13} +\\\nb
&& r ( 4 r{A_1}^{2}{B_1}^{2}{b_1}^{3}b'_1 B'_1+4 {B_1}^{2}A_1 rb_1 b'_1 A'_1+2 r{B_1}^{2}{A'_1}^{2}{b_1}^{2}+\\\nb
&&2 {B_1}^{2}A_1 r{b_1}^{4}B'_1 A'_1+4 A_1 r{B_1}^{3}{b_1}^{3}b'_1 A'_1+{A'_1}^{2}B_1 r+r{A_1}^{2}B_1 {b_1}^{4}{B'_1}^{2}+\\\nb
&&4 r{A_1}^{2}{B_1}^{3}{b_1}^{2}{b'_1}^{2}+2 A_1 rB_1 {b_1}^{2}B'_1 A'_1+r{B_1}^{3}{A'_1}^{2}{b_1}^{4} )\, c_{14} +\\\nb
&& r ( 4 A_1 r{B_1}^{3}{b_1}^{4}A''_1+8 b_1 r{A_1}^{2}B_1 b''_1 B'_1+2 A_1 rB_1 {b_1}^{2}B'_1 A'_1+\\\nb
&&8 r{A_1}^{2}{B_1}^{2}{b_1}^{3}b'_1 B'_1+2 {B_1}^{2}A_1 r{b_1}^{4}B'_1 A'_1+8 r{A_1}^{2}{B_1}^{3}{b'_1} {b_1}^{3}+\\\nb
&&8 A_1 r{B_1}^{3}{b_1}^{3}b'_1 A'_1+4 {b_1}^{4}r{A_1}^{2}{B_1}^{2}B''_1+4 r{A_1}^{2}{B_1}^{3}{b_1}^{2}{b'_1}^{2}+\\\nb
&&8 b_1 r{A_1}^{2}{B_1}^{2}{b'_1}-3 r{A_1}^{2}B_1 {b_1}^{4}{B'_1}^{2}+4 A''_1 {B_1}^{2}rA_1 {b_1}^{2}+\\\nb
&&8 {B_1}^{2}A_1 rb_1 b'_1 A'_1+8 {b_1}^{4}A_1 {B_1}^{3}A'_1-3 r{B_1}^{2}{A'_1}^{2}{b_1}^{2}+4 {b'_1}^{2}r{A_1}^{2}{B_1}^{2}-\\\nb
&&3 {b_1}^{2}{B'_1}^{2}r{A_1}^{2}-3 r{B_1}^{3}{A'_1}^{2}{b_1}^{4}+32 {A_1}^{2}{B_1}^{3}{b_1}^{3}b'_1+8 {B_1}^{2}{A_1}^{2}{b_1}^{4}B'_1+\\\nb
&&32 b_1 {A_1}^{2}{B_1}^{2}b'_1+8 {b_1}^{2}{A_1}^{2}B_1 B'_1+8 {B_1}^{2}A_1 A'_1 {b_1}^{2}+4 {b_1}^{2}r{A_1}^{2}B_1 B''_1 )\, c_{2} +\\\nb
&&  r ( 2 A_1 rB_1 {b_1}^{2}B'_1 A'_1-4 A''_1 B_1 rA_1+4 {b_1}^{2}{A_1}^{2}B_1 B'_1-4 {B_1}^{2}A_1 A'_1 {b_1}^{2}+\\\nb
&&2 r{B_1}^{2}{A'_1}^{2}{b_1}^{2}-4 A''_1 {B_1}^{2}rA_1 {b_1}^{2}+2 B'_1 A'_1 rA_1+2 {A'_1}^{2}B_1 r-\\\nb
&&4 A'_1 B_1 A_1+4 B'_1 {A_1}^{2} )\,],  \\
G^{aether}_{tt}&&=
\frac{1}{8{\dot a_2}^{2}A_1  \left( {b_1}^{2}B_1+1 \right) {B_1}^{2}{r}^{2}}\times\\\nb
&&[\,  ( -3 {b_1}^{2}{B'_1}^{2}{r}^{2}{A_1}^{2}-3 {r}^{2}{B_1}^{3}{b_1}^{4}{A'_1}^{2}-3 {b_1}^{2}{A'_1}^{2}{r}^{2}{B_1}^{2}+4 {b'_1}^{2}{r}^{2}{A_1}^{2}{B_1}^{2}+\\\nb
&& 8 {b_1}^{3}{r}^{2}A_1 {B_1}^{3}A'_1 b'_1+8 b_1 {r}^{2}A_1 {B_1}^{2}b'_1 A'_1+2 {b_1}^{2}{r}^{2}A_1 B_1 A'_1 B'_1+\\\nb
&& 8 b_1 {r}^{2}{A_1}^{2}B_1 b'_1 B'_1+8 {b_1}^{3}{r}^{2}{A_1}^{2}{B_1}^{2}b'_1 B'_1+2 {b_1}^{4}{r}^{2}A_1 {B_1}^{2}A'_1 B'_1-\\\nb
&&8 {A_1}^{2}{B_1}^{3}{b_1}^{4}-8 {b_1}^{2}{A_1}^{2}{B_1}^{2}-3 {r}^{2}{A_1}^{2}B_1 {b_1}^{4}{B'_1}^{2}+4 {b_1}^{2}{r}^{2}{A_1}^{2}{B_1}^{3}{b'_1}^{2}+\\\nb
&& 16 b_1 r{A_1}^{2}{B_1}^{2}b'_1+8 {b_1}^{2}r{A_1}^{2}B_1 B'_1+8 {B_1}^{2}A_1 rA'_1 {b_1}^{2}+8 {b_1}^{4}A_1 {B_1}^{3}A'_1 r+\\\nb
&&16 r{A_1}^{2}{B_1}^{3}{b_1}^{3}b'_1+8 {b_1}^{4}r{A_1}^{2}{B_1}^{2}B'_1+4 A_1 {r}^{2}{B_1}^{3}A''_1 {b_1}^{4}+\\\nb
&& 4 {b_1}^{2}{r}^{2}A_1 {B_1}^{2}A''_1+4 {r}^{2}B_1 {A_1}^{2}{b_1}^{2}B''_1+4 {r}^{2}{B_1}^{2}{A_1}^{2}{b_1}^{4}B''_1+\\\nb
&&8 {r}^{2}{B_1}^{2}{A_1}^{2}b_1 {b''_1}+8 {r}^{2}{B_1}^{3}{A_1}^{2}{b''_1} {b_1}^{3} )\, c_{13} +\\\nb
&&  ( 2 {b_1}^{2}{B'_1}^{2}{r}^{2}{A_1}^{2}+3 {r}^{2}{B_1}^{3}{b_1}^{4}{A'_1}^{2}+6 {b_1}^{2}{A'_1}^{2}{r}^{2}{B_1}^{2}-8 {b'_1}^{2}{r}^{2}{A_1}^{2}{B_1}^{2}-\\\nb
&& 8 A_1 rA'_1 B_1-8 {b_1}^{3}{r}^{2}A_1 {B_1}^{3}A'_1 b'_1-8 b_1 {r}^{2}A_1 {B_1}^{2}b'_1 A'_1-\\\nb
&& 12 b_1 {r}^{2}{A_1}^{2}B_1 b'_1 B'_1-8 {b_1}^{3}{r}^{2}{A_1}^{2}{B_1}^{2}b'_1 B'_1-2 {b_1}^{4}{r}^{2}A_1 {B_1}^{2}A'_1 B'_1+\\\nb
&&3 {r}^{2}{A_1}^{2}B_1 {b_1}^{4}{B'_1}^{2}-4 {b_1}^{2}{r}^{2}{A_1}^{2}{B_1}^{3}{b'_1}^{2}-16 b_1 r{A_1}^{2}{B_1}^{2}b'_1-\\\nb
&&8 {b_1}^{2}r{A_1}^{2}B_1 B'_1-16 {B_1}^{2}A_1 rA'_1 {b_1}^{2}-8 {b_1}^{4}A_1 {B_1}^{3}A'_1 r-\\\nb
&&16 r{A_1}^{2}{B_1}^{3}{b_1}^{3}b'_1-8 {b_1}^{4}r{A_1}^{2}{B_1}^{2}B'_1+3 {A'_1}^{2}B_1 {r}^{2}-4 A_1 {r}^{2}{B_1}^{3}A''_1 {b_1}^{4}-\\\nb
&&8 {b_1}^{2}{r}^{2}A_1 {B_1}^{2}A''_1-4 {r}^{2}B_1 {A_1}^{2}{b_1}^{2}B''_1-4 {r}^{2}{B_1}^{2}{A_1}^{2}{b_1}^{4}B''_1-\\\nb
&& 8 {r}^{2}{B_1}^{2}{A_1}^{2}b_1 {b''_1}-8 {r}^{2}{B_1}^{3}{A_1}^{2}{b'_1} {b_1}^{3}+2 B'_1 A'_1 {r}^{2}A_1-4 A''_1 B_1 {r}^{2}A_1 )\, c_{14} +\\\nb
&&  ( 32 r{A_1}^{2}{B_1}^{3}{b_1}^{3}b'_1+8 {b_1}^{4}r{A_1}^{2}{B_1}^{2}B'_1-3 {r}^{2}{A_1}^{2}B_1 {b_1}^{4}{B'_1}^{2}+\\\nb
&&8 b_1 {r}^{2}{A_1}^{2}B_1 b'_1 B'_1+2 {b_1}^{2}{r}^{2}A_1 B_1 A'_1 B'_1+4 {b_1}^{2}{r}^{2}{A_1}^{2}{B_1}^{3}{b'_1}^{2}+\\\nb
&&8 b_1 {r}^{2}A_1 {B_1}^{2}b'_1 A'_1+2 {b_1}^{4}{r}^{2}A_1 {B_1}^{2}A'_1 B'_1+32 b_1 r{A_1}^{2}{B_1}^{2}b'_1+\\\nb
&&8 {b_1}^{2}r{A_1}^{2}B_1 B'_1+8 {B_1}^{2}A_1 rA'_1 {b_1}^{2}+8 {b_1}^{4}A_1 {B_1}^{3}A'_1 r+8 {r}^{2}{B_1}^{2}{A_1}^{2}b_1 {b''_1}+\\\nb
&&4 {r}^{2}B_1 {A_1}^{2}{b_1}^{2}B''_1+4 {b_1}^{2}{r}^{2}A_1 {B_1}^{2}A''_1+4 A_1 {r}^{2}{B_1}^{3}A''_1 {b_1}^{4}+\\\nb
&&8 {r}^{2}{B_1}^{3}{A_1}^{2}{b''_1} {b_1}^{3}+4 {r}^{2}{B_1}^{2}{A_1}^{2}{b_1}^{4}B''_1+8 {b_1}^{3}{r}^{2}A_1 {B_1}^{3}A'_1 b'_1+\\\nb
&&8 {b_1}^{3}{r}^{2}{A_1}^{2}{B_1}^{2}b'_1 B'_1-3 {b_1}^{2}{A'_1}^{2}{r}^{2}{B_1}^{2}-3 {b_1}^{2}{B'_1}^{2}{r}^{2}{A_1}^{2}-\\\nb
&& 3 {r}^{2}{B_1}^{3}{b_1}^{4}{A'_1}^{2}+4 {b'_1}^{2}{r}^{2}{A_1}^{2}{B_1}^{2} )\, c_{2}+\\\nb
&& 8 {A_1}^{2}{B_1}^{3}{b_1}^{2}+8 {B_1}^{2}{A_1}^{2}-8 B_1 {A_1}^{2}-8 {b_1}^{2}{A_1}^{2}{B_1}^{2}+\\\nb
&&8 {b_1}^{2}r{A_1}^{2}B_1 B'_1+8 rB'_1 {A_1}^{2}\, ],\\
G^{aether}_{tr}&&=
-\frac{1}{4{r}^{2}{B_1}^{2}\dot a_2  \left[ A_1  \left( {b_1}^{2}B_1+1 \right)  \right] ^{\frac{3}{2}}}  \times\\\nb
&&[\, ( -8 {B_1}^{2}{A_1}^{2}b_1-8 {A_1}^{2}{B_1}^{4}{b_1}^{5}-16 {B_1}^{3}{A_1}^{2}{b_1}^{3}+2 {B_1}^{2}{r}^{2}A_1 A'_1 b'_1+\nb\\
&& 2 B_1 {r}^{2}{A_1}^{2}b'_1 B'_1+8 r{A_1}^{2}{B_1}^{4}{b_1}^{4}b'_1+16 {B_1}^{3}r{A_1}^{2}{b_1}^{2}b'_1+\\\nb
&& 8 {B_1}^{3}rA_1 {b_1}^{3}A'_1+4 {B_1}^{2}rA_1 b_1 A'_1+4 {b_1}^{5}A_1 {B_1}^{4}rA'_1+4 {B_1}^{3}r{A_1}^{2}{b_1}^{5}B'_1-\\\nb
&&2 {B_1}^{2}{r}^{2}{A_1}^{2}{b_1}^{5}{B'_1}^{2}-4 B_1 {r}^{2}{A_1}^{2}{b_1}^{3}{B'_1}^{2}+8 {B_1}^{2}r{A_1}^{2}{b_1}^{3}B'_1+\\\nb
&& 4 B_1 r{A_1}^{2}b_1 B'_1+2 {B_1}^{3}{r}^{2}{A_1}^{2}{b_1}^{5}B''_1+4 {B_1}^{2}{r}^{2}{A_1}^{2}{b_1}^{3}B''_1+\\\nb
&& 2 B_1 {r}^{2}{A_1}^{2}b_1 B''_1+4 {r}^{2}{B_1}^{4}{A_1}^{2}{b''_1} {b_1}^{4}+8 {B_1}^{3}{r}^{2}{A_1}^{2}{b_1}^{2}{b''_1}+\\\nb
&& 2 A_1 {B_1}^{4}{r}^{2}{b_1}^{5}A''_1+4 {B_1}^{3}{r}^{2}A_1 {b_1}^{3}A''_1+2 {B_1}^{2}{r}^{2}A_1 b_1 A''_1+\\\nb
&& 4 A'_1 {B_1}^{3}{r}^{2}A_1 b'_1 {b_1}^{2}+2 A_1 {B_1}^{4}{r}^{2}A'_1 {b_1}^{4}b'_1+4 {B_1}^{2}{r}^{2}{A_1}^{2}b'_1 B'_1 {b_1}^{2}+\\\nb
&& 2 {B_1}^{3}{r}^{2}{A_1}^{2}{b_1}^{4}b'_1 B'_1+8 {B_1}^{2}r{A_1}^{2}b'_1+4 {B_1}^{2}{r}^{2}{A_1}^{2}{b'_1}-2 {B'_1}^{2}{r}^{2}{A_1}^{2}b_1-\\\nb
&& 2 {r}^{2}{B_1}^{4}{b_1}^{5}{A'_1}^{2}-4 {A'_1}^{2}{B_1}^{3}{r}^{2}{b_1}^{3}-2 {A'_1}^{2}{B_1}^{2}{r}^{2}b_1 )\, c_{13} +\\\nb
&& ( -8 r{A_1}^{2}{B_1}^{4}{b_1}^{4}b'_1-8 {B_1}^{3}r{A_1}^{2}{b_1}^{2}b'_1-8 {B_1}^{3}rA_1 {b_1}^{3}A'_1-\\\nb
&& 4 {B_1}^{2}rA_1 b_1 A'_1-4 {b_1}^{5}A_1 {B_1}^{4}rA'_1-4 {B_1}^{3}r{A_1}^{2}{b_1}^{5}B'_1+\\\nb
&& 2 {B_1}^{2}{r}^{2}{A_1}^{2}{b_1}^{5}{B'_1}^{2}+B_1 {r}^{2}{A_1}^{2}{b_1}^{3}{B'_1}^{2}-4 {B_1}^{2}r{A_1}^{2}{b_1}^{3}B'_1-\\\nb
&& 2 {B_1}^{3}{r}^{2}{A_1}^{2}{b_1}^{5}B''_1-2 {B_1}^{2}{r}^{2}{A_1}^{2}{b_1}^{3}B''_1-4 {r}^{2}{B_1}^{4}{A_1}^{2}{b''_1} {b_1}^{4}-\\\nb
&& 4 {B_1}^{3}{r}^{2}{A_1}^{2}{b_1}^{2}{b''_1}-2 A_1 {B_1}^{4}{r}^{2}{b_1}^{5}A''_1-4 {B_1}^{3}{r}^{2}A_1 {b_1}^{3}A''_1-\\\nb
&& 2 {B_1}^{2}{r}^{2}A_1 b_1 A''_1-4 {B_1}^{3}{r}^{2}{A_1}^{2}b_1 {b'_1}^{2}-2 A'_1 {B_1}^{3}{r}^{2}A_1 b'_1 {b_1}^{2}-\\\nb
&& 2 A_1 {B_1}^{4}{r}^{2}A'_1 {b_1}^{4}b'_1-6 {B_1}^{2}{r}^{2}{A_1}^{2}b'_1 B'_1 {b_1}^{2}-2 {B_1}^{3}{r}^{2}{A_1}^{2}{b_1}^{4}b'_1 B'_1+\\\nb
&& B_1 {r}^{2}A_1 b_1 A'_1 B'_1+{B_1}^{2}{r}^{2}A_1 A'_1 {b_1}^{3}B'_1+2 {r}^{2}{B_1}^{4}{b_1}^{5}{A'_1}^{2}+\\\nb
&& 4 {A'_1}^{2}{B_1}^{3}{r}^{2}{b_1}^{3}+ 2 {A'_1}^{2}{B_1}^{2}{r}^{2}b_1 )\, c_{14} +\\\nb
&& ( -8 {B_1}^{2}{A_1}^{2}b_1-8 {A_1}^{2}{B_1}^{4}{b_1}^{5}-16 {B_1}^{3}{A_1}^{2}{b_1}^{3}+2 {B_1}^{2}{r}^{2}A_1 A'_1 b'_1+\\\nb
&& 2 B_1 {r}^{2}{A_1}^{2}b'_1 B'_1+8 r{A_1}^{2}{B_1}^{4}{b_1}^{4}b'_1+16 {B_1}^{3}r{A_1}^{2}{b_1}^{2}b'_1-\\\nb
&& 2 {B_1}^{2}{r}^{2}{A_1}^{2}{b_1}^{5}{B'_1}^{2}-4 B_1 {r}^{2}{A_1}^{2}{b_1}^{3}{B'_1}^{2}+2 {B_1}^{3}{r}^{2}{A_1}^{2}{b_1}^{5}B''_1+\\\nb
&& 4 {B_1}^{2}{r}^{2}{A_1}^{2}{b_1}^{3}B''_1+2 B_1 {r}^{2}{A_1}^{2}b_1 B''_1+4 {r}^{2}{B_1}^{4}{A_1}^{2}{b''_1} {b_1}^{4}+\\\nb
&& 8 {B_1}^{3}{r}^{2}{A_1}^{2}{b_1}^{2}{b''_1}+2 A_1 {B_1}^{4}{r}^{2}{b_1}^{5}A''_1+4 {B_1}^{3}{r}^{2}A_1 {b_1}^{3}A''_1+\\\nb
&& 2 {B_1}^{2}{r}^{2}A_1 b_1 A''_1+4 A'_1 {B_1}^{3}{r}^{2}A_1 b'_1 {b_1}^{2}+2 A_1 {B_1}^{4}{r}^{2}A'_1 {b_1}^{4}b'_1+\\\nb
&& 4 {B_1}^{2}{r}^{2}{A_1}^{2}b'_1 B'_1 {b_1}^{2}+2 {B_1}^{3}{r}^{2}{A_1}^{2}{b_1}^{4}b'_1 B'_1+8 {B_1}^{2}r{A_1}^{2}b'_1+\\\nb
&& 4 {B_1}^{2}{r}^{2}{A_1}^{2}{b''_1}-2 {B'_1}^{2}{r}^{2}{A_1}^{2}b_1-2 {r}^{2}{B_1}^{4}{b_1}^{5}{A'_1}^{2}-4 {A'_1}^{2}{B_1}^{3}{r}^{2}{b_1}^{3}-\\\nb
&& 2 {A'_1}^{2}{B_1}^{2}{r}^{2}b_1 )\, c_{2} \,],\\
\lb{Gaetr1}
\eqn
where $A_1=({ b_1^2 B_1+1})/({a_1^2 a_2^2})$.
We can notice that solving these field equations gives precisely the same results as our previous paper \cite{Chan2022}. There, we have shown that all the solutions with specific values of $c_2$, $c_1+c_3$, and $c_1+c_4$ are static black holes with universal horizons that are asymptotically flat. These two conditions must be fulfilled for the BT to be obeyed. However, we cannot say anything about the uniqueness of the solutions because of the complexity of the field equations.

\section{Conclusions}

We set out to investigate the BT in EA theory, a Lorentz violating the theory of gravity. We found that the answer to the question posed in the title is  `Yes,' when aether has only a temporal component and for very particular values of the constants of the theory. However, when the aether has both temporal and radial components, we have shown that the staticity and flatness at infinity hold only for a particular metric suggested by the Killing vectors. Besides, for both cases, only for particular values of $c_2$, $c_1+c_3$ and $c_1+c_4$ are provable. In the first case, these solutions violate cosmic censorship conjecture \cite{Chan2020}; however, in the second case, these solutions have universal horizons \cite{Chan2022}.

The reader can see that the metric employed in both cases is the most general spherically symmetric metric. However, the aether vector in equation (\ref{uar}) is more general than the one in equation (\ref{uat}) where $b(r,t)=0$.	 However, the latter case has a problem. The aether vector with only a temporal component gives a timelike Killing vector that does not become null at the horizon. This prevents the solution from being a black hole. In other words, we end up with a naked singularity. 

From the analysis of Killing vectors, we show that, in both cases, spacetime has three spacelike Killing vectors satisfying the Lie algebra of the SO(3) group of isometries. It shows that our metric has a spherical symmetry. However, the fourth timelike Killing vector only exists when $B(r,t)$ and $b(r,t)$ are purely radial functions, and $a(r,t)$ is either purely radial or it can be written as a variable separable function. This will automatically render $A(r,t)$ to either purely radial or variable separable function. This means that to prove the BT in EA theory for an aether vector with time and radial components, we have to use a particular metric suggested by the timelike Killing vector. These solutions have universal horizons. While we show the existence of BT in EA theory for the case where the aether vector has only a time component, the solutions turn out to be naked singularities and violate cosmic censorship conjecture. 

\section {Acknowledgments}

The author (RC) acknowledges the financial support from FAPERJ (no.E-26/171.754/2000, E-26/171.533/2002 and E-26/170.951/2006). MFAdaS acknowledges the financial support from CNPq-Brazil, FINEP-Brazil (Ref. 2399/03), FAPERJ/UERJ (307935/2018-3), { FAPERJ (E-26/211.906/2021)} and from CAPES (CAPES-PRINT 41/2017). 

\section{References}

\end{document}